**Spatial distribution of nuclei in progressive nucleation: modeling and application**

Massimo Tomellini

Dipartimento di Scienze e Tecnologie Chimiche, Università degli Studi di Roma Tor Vergata

Via della Ricerca Scientifica 00133 Roma Italy

**Abstract**

Phase transformations ruled by non-simultaneous nucleation and growth do not lead to random distribution of nuclei. Since nucleation is only allowed in the untransformed portion of space, positions of nuclei are correlated. In this article an analytical approach is presented for computing pair-correlation function of nuclei in progressive nucleation. This quantity is further employed for characterizing the spatial distribution of nuclei through the nearest neighbor distribution function. The modeling is developed for nucleation in 2D space with power growth law and it is applied to describe electrochemical nucleation where correlation effects are significant. Comparison with both computer simulations and experimental data lends support to the model which gives insights into the transition from Poissonian to correlated nearest neighbor probability density.

## 1-Introduction

Phase transformations ruled by nucleation and growth represents an important topic in Materials Science because of the effect these processes have on the microscopic structure of the materials. In modeling phase transformations, nucleation is assumed to occur at seeds, randomly distributed throughout the space, which become nuclei once they start growing. Nucleation can be either simultaneous or progressive; in the first case nuclei are all formed at the same instant while in the second are generated continuously as the transformation proceeds. The transformation can be analysed by the theory of nucleation and growth developed, independently, by Kolmogorov, Johnson and Mehl and Avrami (KJMA) [1-3]. Progressive nucleation rises some issues at the level of modeling; the most celebrated one is related to the "phantom overgrowth" which limits the applicability of the KJMA approach to a particular class of growth laws [4]. The term phantom, as originally introduced in ref.[3], is referred to a seed that was swallowed by the new phase before it starts growing. Such a seed will not contribute to the phase transformation. Nevertheless, in the framework of Poisson dot process both actual and phantom nuclei have to be included in the mathematical formulation, since seeds are randomly distributed throughout the space [4]. The presence of phantom nuclei put in evidence another difference between simultaneous and progressive nucleation, that is related to the spatial distribution of actual nuclei. In fact, from the previous considerations it follows that in the former case nuclei are randomly distributed throughout the space while in the latter non-Poissonian distribution is expected.

On one hand several studies have been devoted to model the *size*-distribution function of nuclei in phase transformations which are compliant with the KJMA approach. These modelings are aimed at determining the time dependence of the distribution function for both simultaneous and progressive nucleation [5-7]. In the case of simultaneous nucleation the distribution function at the end of the transformation matches the Gamma distribution which well describes Poisson-Voronoi tessellation even in non-Euclidean metrics [5,8,9]. Progressive nucleation is more involved as it implies fragmentation of Voronoi cells during nucleation; yet a good description of the distribution function has been achieved through superposition of Gamma distribution functions [7]. Insights into macroscopic properties of materials can also be gained through the study of correlation functions among transformed and/or untransformed points of the system [10-12]. On the other hand, modelings devoted to characterize the spatial distribution of the actual nuclei in KJMA-type phase transformations are scarce in the literature. This is due to the fact that a comprehensive description of the microscopic structure of the system is achieved by means of the quantities above mentioned, linked to size and shape of the grains of the product phase. Nevertheless, the study of spatial arrangement of actual nuclei in KJMA compliant transformations ruled by progressive nucleation is of great interest in view of its application to electrodeposition. At large overpotentials the diffusion of the active species in the liquid phase becomes rate determining and diffusion fields are established around growing nuclei. As the deposition proceeds, overlap among diffusion fields leads the transition from a spherical to a planar regime of diffusion [13-14]. In order to describe this complex transition, modeling based on the concept of " diffusion zones" has been developed which is suitable for describing experimental kinetics [15-18]. This approach exploits the analogy between the diffusion current at the surface of a hemispherical nucleus and that at a planar surface [19]. The complex problem of describing mass transport for overlapping hemispherical diffusion fields is therefore reduced to study the planar diffusion across overlapping 2D disks [13,14]. In the planar diffusion zones approach a 1:1 correspondence is assumed between actual nuclei and 2D disks, the growth of each nucleus is modeled by diffusion through its own disk (or a part of it). It follows that during progressive nucleation an exclusion zone for nucleation develops around each actual nucleus [13]. The radius of the exclusion disk ($r_d$) is computed to be greater than the nucleus radius ($r_n$) [14] where the area of the diffusion zone is computed by means of the KJMA model. Owing to the large $r_d/r_n$ ratio, when nucleation is exhausted the fraction of electrode surface covered by nuclei is small, nuclei are well separated and their distribution is, in general, non-Poissonian [20,21]. It is also in this ambit that the modeling of the distribution of actual nuclei in KJMA-type transitions finds its justification.

Characterization of the spatial distribution of actual nuclei in electrodeposition is performed through the nearest neighbor probability density and pair distribution functions. Experimental data on a variety of electrochemical systems [22-25] are successfully interpreted on the basis of the "exclusion zone" model for nucleation. Computer simulations also show that the spatial arrangement is ruled by the most influential neighbor, so supporting the exclusion zone hypothesis above reported [26]. Studies on the possibility to get long-range order in electrodeposition have been carried out in ref.[24] and distribution function for n[th]-neighbors determined by computer simulations in ref.[27]. On one hand, computer simulations of progressive nucleation with exclusion zones show that the displacement of the nearest neighbor distribution (nnd) from the Poissonian distribution depends upon number density of nucleation sites [21,27]. On the other hand, analytical approaches of the nnd in electrodeposition are limited to the hard-core correlation between disks equal in size [20].

The purpose of the present work is twofold: Firstly, to develop an analytical model of the pair-correlation function and nearest neighbor probability density of actual nuclei in KJMA transformations with progressive nucleation. Secondly, to apply the modeling for describing nnd in electrodeposition and to compare it with computer simulations and experimental data. In view of this application the theory is developed for transformation in 2D space, although its extension to other dimensions is straightforward.

The paper is organized as follows. The first two sections are devoted to the computation of the pair-correlation function and nearest neighbor probability density. The purpose of the third section is to bridge the gap between computer simulations and analytical approach for nnd in electrochemical nucleation. To this end, the last section provides application of the model to describe nnd obtained from experiments and computer simulations.

## 2-Results and discussion

### 2.1 Pair-correlation function of actual nuclei

In this section we determine the pair-correlation function of actual nuclei in 2D transformations occurring by progressive nucleation and growth. Throughout the paper, we distinguish between actual and phantom (or virtual) nuclei since they are both considered in the formulation of the theory. As anticipated in the introduction, the present work also focuses on modeling the spatial distribution of actual nuclei in electrodeposition. To this end, upon nucleation a disk of radius $r_d$, centered on each nucleus starts growing, that is the (time dependent) disk where further nucleation is prevented. In the following the term "exclusion zone" is referred to the region unavailable to the formation of actual nuclei. The area of the exclusion zone is computed through the KJMA theory for nucleation and growth of disks of radius $r_d$, where $r_d$ is greater than nucleus radius. Clearly, if the radius of the exclusion disk coincides with the nucleus radius the "exclusion zone" is the "natural region" where actual nuclei do not form, being this region already transformed by the new phase. In other words, the exclusion zone coincides with the transformed phase.

The pair-correlation function is linked to the probability of finding two actual nuclei in volume elements $d\boldsymbol{r}_1$ and $d\boldsymbol{r}_2$ located, respectively, at $\boldsymbol{r}_1$ and $\boldsymbol{r}_2$, independently of the location of the other nuclei. By denoting with $dP(\boldsymbol{r}_1, \boldsymbol{r}_2)$ this probability, we get $dP(\boldsymbol{r}_1, \boldsymbol{r}_2) = f(\boldsymbol{r}_1, \boldsymbol{r}_2)d\boldsymbol{r}_1 d\boldsymbol{r}_2$ where $f(\boldsymbol{r}_1, \boldsymbol{r}_2)$ is the two-particle distribution function also called "$f$-function" [28]. Moreover, $f$ is linked to the relative density at $\boldsymbol{r}_2$, $f_c(\boldsymbol{r}_1|\boldsymbol{r}_2)$, through the relation $f(\boldsymbol{r}_1, \boldsymbol{r}_2) = N(\boldsymbol{r}_1)f_c(\boldsymbol{r}_1|\boldsymbol{r}_2)$ where $N(\boldsymbol{r}_1)$ is the density of nuclei at $\boldsymbol{r}_1$. The pair-correlation function is defined as $g(\boldsymbol{r}_1, \boldsymbol{r}_2) = \frac{f(\boldsymbol{r}_1, \boldsymbol{r}_2)}{N(\boldsymbol{r}_1)N(\boldsymbol{r}_2)} = \frac{f_c(\boldsymbol{r}_1|\boldsymbol{r}_2)}{N(\boldsymbol{r}_2)}$, that is $f_c(\boldsymbol{r}_1|\boldsymbol{r}_2) = N(\boldsymbol{r}_2)g(\boldsymbol{r}_1, \boldsymbol{r}_2)$. If the system is homogeneous and isotropic, $N(\boldsymbol{r}_1) = N$, that is independent of spatial coordinates. Under these circumstances $f_c = f_c(|\boldsymbol{r}_2 - \boldsymbol{r}_1|) = f_c(r)$, that is $g(r) = \frac{f(r)}{N^2}$. Since $f_c(r)d\boldsymbol{r}$ is equal to the number of nuclei at relative distance $r$ from the one located at $\boldsymbol{r}_1$, it follows that the local density is given by $\rho(r) = Ng(r)$. In the following, we determine the radial distribution function, $g(r)$, by exploiting its definition in terms of *f*-function.

In the framework of the KJMA model, in order to estimate the *f*-function we consider nucleation to occur randomly throughout the *entire* surface. A nucleus (either actual or phantom) is generated in time-space volume element $d\boldsymbol{r}dt'$ centered at $(\boldsymbol{r}, t')$ with probability equal to $I(t')d\boldsymbol{r}dt'$, with $I(t')$ being the nucleation rate per unit surface (phantom-included). In the following, the actual time (time of observation) will be denoted with $t$ and the radius of the exclusion disk, for a seed which transforms in time interval $dt'$ at $t'$, as $R(t - t')$. In the computation that follows *t* enters as a parameter which determines the fraction of surface prevented to the formation of actual nuclei.

Let us consider two points of the space located at $\boldsymbol{r}_1$ and $\boldsymbol{r}_2$ at time $t'$ and $t''$, respectively, with $t' < t'' < t$ (Fig. 1). The probability that two actual nuclei are generated within $d\boldsymbol{r}_1 dt'$ and $d\boldsymbol{r}_2 dt''$ at $(\boldsymbol{r}_1, t')$ and $(\boldsymbol{r}_2, t'')$, respectively, reads

$$F(t', t'', r)dt'dt''d\boldsymbol{r}_1 d\boldsymbol{r}_2 = [P_1(t')I(t')dt'd\boldsymbol{r}_1][P_2(t', t'', r)I(t'')dt''d\boldsymbol{r}_2], \tag{1}$$

where $F$ is the probability density and $r = |\boldsymbol{r}_2 - \boldsymbol{r}_1|$ is subjected to the constraint $r > R(t'' - t')$[1]. Similar equation holds for $t'' < t'$. In the second member of eqn.1a, the term in the first brackets is the probability that an actual nucleus is generated within $d\boldsymbol{r}_1 dt'$ at $(\boldsymbol{r}_1, t')$; the last term in the square brackets is the probability that an actual nucleus is formed within $d\boldsymbol{r}_2 dt''$ at $(\boldsymbol{r}_2, t'')$, given that an actual nucleus has already been formed at $(\boldsymbol{r}_1, t')$. The probabilities $P_1(t')$ and $P_2(t', t'', r)$ are given by:

$$P_1(t') = \exp\left[-\int_0^{t'} A_1(t', \tau)I(\tau)\, d\tau\right] \tag{2a}$$

---

[1] This condition ensures that at time $t''$ the second seed is not covered by the first nucleus; consequently a phantom does not nucleate at $\boldsymbol{r}_2$ between $t''$ and $t'' + dt''$.

$$P_2(t',t'',r) = \exp[-\int_0^{t'} A_2(t',t'',\tau,r)I(\tau)\,d\tau + \int_{t'}^{t''} A_3(t',t'',\tau,r)I(\tau)\,d\tau] \qquad (2b)$$

where

$$A_1(t',\tau) = \pi R^2(t'-\tau) \qquad (3a)$$

$$A_2(t',t'',\tau,r) = \pi R^2(t''-\tau) - \omega_1[R(t'-\tau),R(t''-\tau);r], \qquad (3b)$$

$$A_3(t',t'',\tau,r) = \pi R^2(t''-\tau) - \omega_2[R(t''-\tau),R(\tau-t');r] \qquad (3c)$$

In eqns.3b-c, $\omega[R_1,R_2;r]$ is the overlap area between two disks of radius $R_1$ and $R_2$ at relative distance $r$. Eqns.1,2 hold true since $I(t)$ is the whole nucleation rate and nuclei (actual and phantoms) are distributed at random throughout the surface. A graphical representation of the regions with areas $A_1$, $A_2$, $A_3$ and $\omega_i$ ($i$=1,2) is shown in Fig.1. Nuclei which start growing within these regions are capable of transforming the points at $\boldsymbol{r}_1$ and $\boldsymbol{r}_2$ before time $t'$ and $t''$, respectively. These regions are defined as the capture zones for the two points [29]. In eqn.3b the overlap area $\omega_1$ gives account of the contribution already included in $A_1$; in the conditional probability eqns.2b, 3c, $\omega_2$ is related to the fact that, for $\tau > t'$, nucleation of phantoms is allowed in the region "transformed" by the first actual nucleus. By setting $P(t',t'',r) = P_1(t')P_2(t',t'',r)$, the probability to get an actual nucleus at $\boldsymbol{r}_2$ within time $t$ and an actual nucleus at $\boldsymbol{r}_1$ which was born between $t'$ and $t'+dt'$, is attained by integrating over $t''$:

$$\tilde{F}(t,t',r)dt'd\boldsymbol{r}_1 d\boldsymbol{r}_2 = I(t')dt'd\boldsymbol{r}_1 d\boldsymbol{r}_2 \ \left[\int_0^{t'} dt'' \, P(t',t'',r)I(t'')H(r-R(t'-t'')) + \int_{t'}^{t} dt'' \, P(t'',t',r)I(t'')H(r-R(t''-t'))\right],$$

$$\qquad (4)$$

where $H(x)$ is the Heaviside function. The $f$-function is eventually computed by integrating $\tilde{F}$ over $t'$ as

$$f(r,t) = \int_0^t I(t')dt' \int_0^{t'} dt'' \, P(t',t'',r)I(t'')H(r-R(t'-t''))$$

$$+ \int_0^t I(t')dt' \int_{t'}^t dt'' \, P(t'',t',r)I(t'')H(r-R(t''-t')). \qquad (5a)$$

Also, changing the order of integration in the second integral one obtains,

$$f(r,t) = \int_0^t dt' \int_0^{t'} dt'' \, P(t',t'',r)I(t')I(t'')H(r-R(t'-t''))$$

$$+ \int_0^t dt'' \int_0^{t''} dt' \, P(t'',t',r)I(t')I(t'')H(r-R(t''-t'))$$

$$= 2 \int_0^t dt'' \int_0^{t''} dt' \, P(t'',t',r)I(t')I(t'')H(r-R(t''-t')). \qquad (5b)$$

This is the equation we employ for computing the $g(r)$ and the nnd function in the case of constant nucleation rate and power growth law, namely $I(t) = I_0$ and $R(t) = (\gamma t)^{n/2}$.

Using dimensionless variables, $x' = t'/t$, $x'' = t''/t$, $\eta = \tau/t$, $\bar{\omega}_i = \omega_i/(\gamma t)^n$, and $\rho = r/(\gamma t)^{n/2}$, the two-nuclei distribution function is computed according to (see Appendix A for details)

$$f(\rho,t) = 2I_0^2 t^2$$

$$\times \int_0^1 e^{-S_{ex}(t)x''^{n+1}} dx'' \int_0^{x''} e^{-S_{ex}(t)\left[x'^{n+1} - \frac{n+1}{\pi}\int_0^{x'}\bar{\omega}_1(x',x'',\eta,\rho)d\eta - \frac{n+1}{\pi}\int_{x'}^{x''}\bar{\omega}_2(x',x'',\eta,\rho)d\eta\right]}$$

$$\times H\left(-(x''-x')^{n/2}\right) dx', \qquad (6)$$

where $S_{ex}(t) = \frac{1}{n+1}\pi I_0 \gamma t^{n+1}$ is the extended surface of the exclusion zone for nucleation, and $\bar{\omega}_i(x',x'',\eta,\rho)$ the overlap area at reduced distance $\rho$ ($i$=1,2). In the following, to simplify the notation in some occurrences the $t$-dependence will be omitted from $S_{ex}(t)$. Since the nucleation rate of the actual nuclei is given by $I_a(t) = I_0(1-S(t)) = I_0 e^{-S_{ex}(t)}$, the density of nuclei is $N(t) = I_0 \int_0^t e^{-S_{ex}(\tau)} d\tau = I_0 t \int_0^1 e^{-S_{ex}(t)\xi^{n+1}} d\xi$ and the radial distribution function is eventually computed from the relation $g(r,t) = \frac{f(r,t)}{N(t)^2}$ according to,

$$g(\rho, S_{ex}) =$$

$$= \frac{2 \int_0^1 e^{-S_{ex}x''^{n+1}} dx'' \int_0^{x''} e^{-S_{ex}x'^{n+1}} e^{\frac{n+1}{\pi}S_{ex}\Omega(x',x'',\rho)} H(x'-x''+\rho^{2/n}) dx'}{\left[\int_0^1 e^{-S_{ex}\xi^{n+1}} d\xi\right]^2}, \qquad (7a)$$

where

$$\Omega(x',x'',\rho) = \left[\int_0^{x'} \overline{\omega}_1(x',x'',\eta,\rho) \, d\eta + \int_{x'}^{x''} \overline{\omega}_2(x',x'',\eta,\rho) \, d\eta\right]. \qquad (7b)$$

It is worth pointing out that in the present computation both nucleation rate and actual time are different from zero. In other words, the "phantom-included" nucleation density, $I_0 t$, is assumed to be always different from zero, i.e. we require a non-vanishing $N(t)$. The limiting case $S_{ex}(t) = 0$ is therefore linked to a $\gamma = 0$, which implies no-exclusion area for nucleation. This case can be envisaged as a progressive nucleation without nucleus growth. Furthermore, the present approach could be employed to study the radial distribution function of a homogeneous system of hard disks equal in size. This should require a simultaneous nucleation with delta-function growth rate according to $R(t) = R_0 H(t)$ [20]. Notably, for this growth law but for progressive nucleation the KJMA model cannot be applied owing to a non-negligible effect of phantom "overgrowth" [30,31].

### 2.2 Nearest-neighbor distribution function

The nearest neighbor distribution function can be computed from the knowledge of the conditional radial distribution function by exploiting the approach presented by Torquato et al in ref.[32]. Accordingly, the nearest- neighbor probability density, $H_p(r)$, is given by

$$H_p(r) = N2\pi r G(r) E_p(r), \qquad (8)$$

where $E_p(r)$ is the probability that, given a circular region of radius $r$ encompassing a particle (in our case an actual nucleus), this region is empty of particle centers (actual nuclei) and $N2\pi r G(r)dr$ is the probability that, given the condition above stated for $E_p(r)$, actual nuclei are contained in the shell $2\pi r dr$ surrounding the central nucleus. Since $E_p(r) = 1 - \int_0^r H_p(r')dr'$ it follows that

$$H_p(r) = -\frac{\partial E_p(r)}{\partial r}. \tag{9}$$

The system of eqns.8-9 provides $H_p(r)$ as a function of $G(r)$:

$$H_p(r) = 2\pi N r G(r) \exp\left[-2\pi N \int_0^r y G(y) dy\right]. \tag{10}$$

In the following, eqn.10 is evaluated by setting $G(r) \approx g(r)$ and the goodness of the approximation tested through comparison with computer simulations. In terms of the dimensionless variable defined above, $\rho$, the nearest-neighbor probability density is given by the relationship $H_p(\rho) = H_p(r(\rho))\frac{dr}{d\rho}$ with $\rho = \frac{r}{(\gamma t)^{n/2}}$. Using eqn.10 in this last expression and recalling the definition of the extended surface, one obtains[2]

$$H_p(\rho, S_{ex}) = 2(n+1)S_{ex}\phi_n(S_{ex})\rho g(\rho, S_{ex}) \exp\left[-2(n+1)S_{ex}\phi_n(S_{ex})\int_0^\rho y g(y, S_{ex})dy\right] \tag{11}$$

where $\phi_n(S_{ex}) = \int_0^1 e^{-S_{ex}\xi^{n+1}}d\xi$ .

For the sake of completeness and in order to make a comparison with Poissonian distribution, it is also profitable to express the $H_p$ probability density in terms of the reduced distance $\tilde{\rho} =$ $r\sqrt{2\pi N(t)}$. Moreover, it is obtained $r = \rho(\gamma t)^{n/2} = \frac{\tilde{\rho}}{\sqrt{2\pi N(t)}}$ which implies $\tilde{\rho} = \rho\sqrt{2(n+1)S_{ex}\phi_n(S_{ex})}$. From the relationship $H_p(\tilde{\rho}) = H_p(\rho(\tilde{\rho}))\frac{d\rho(\tilde{\rho})}{d\tilde{\rho}}$ the nearest neighbor distribution becomes:

$$H_p(\tilde{\rho}) = \tilde{\rho}\, g(\tilde{\rho}, S_{ex}) \exp\left[-2(n+1)S_{ex}\phi_n(S_{ex})\int_0^{\rho(\tilde{\rho})} y g(y, S_{ex}) dy\right]. \tag{12}$$

---

[2] To avoid multiple notation the same symbols, $H_p$ and $g$, are retained independently of reduced distance variable.

In the case of random distribution of nuclei $g(r) = 1$ and eqn.12 provides the expected result

$$H_p(\tilde{\rho}) = \tilde{\rho}\,\exp[-\tilde{\rho}^2/2]. \tag{13}$$

## 2.3 Numerical results and application to electrodeposition

In view of the application to electrodeposition, in this section we report the outcomes of the computation of both $g$ and $H_p$ functions as given by eqns.7,11,12 in the case of diffusional growth, i.e. for $n = 1$. We point out that diffusional growth is not compliant with KJMA theory owing to phantom overgrowth. Nevertheless, even in this important case the model can be used as a very good approximation because deviation of the KJMA theory from the exact kinetics are less than 1% [33-35]. For $n$=1 eqns.7a, 12 read

$$g(\rho, S_{ex}) = \frac{2\int_0^1 e^{-S_{ex}x''^2}dx''\int_0^{x''} e^{-S_{ex}x'^2}e^{\frac{2}{\pi}S_{ex}\Omega(x',x'',\rho)}H(x'-x''+\rho^2)dx'}{\left[\int_0^1 e^{-S_{ex}\xi^2}d\xi\right]^2} \tag{14}$$

$$H_p(\tilde{\rho}) = \tilde{\rho}\, g(\tilde{\rho}, S_{ex}) \exp\left[-4S_{ex}\phi_1(S_{ex})\int_0^{\rho(\tilde{\rho})} y g(y, S_{ex}) dy\right], \tag{15}$$

where $\Omega(x', x'', \rho)$ is given by eqn.7b with

$\overline{\omega}_1(x', x'', \eta, \rho) =$

$(x''-\eta)\cos^{-1}\frac{\rho^2-(x'-x'')}{2\rho\sqrt{x''-\eta}} + (x'-\eta)\cos^{-1}\frac{\rho^2+(x'-x'')}{2\rho\sqrt{x'-\eta}} - \frac{1}{2}\sqrt{2\rho^2(x'+x''-2\eta)-\rho^4-(x'-x'')^2}$

and

$$\overline{\omega}_2(x', x'', \eta, \rho) =$$

$$(x'' - \eta) \cos^{-1} \frac{\rho^2 + (x' + x'' - 2\eta)}{2\rho\sqrt{x'' - \eta}} + (\eta - x') \cos^{-1} \frac{\rho^2 - (x' + x'' - 2\eta)}{2\rho\sqrt{\eta - x'}} -$$

$$\frac{1}{2}\sqrt{2\rho^2(x'' - x') - \rho^4 - (2\eta - x' - x'')^2} \ .$$

Also, in eqn.15 $\phi_1(S_{ex}) = \frac{1}{\sqrt{S_{ex}}} \frac{\sqrt{\pi}}{2} \text{erf}(\sqrt{S_{ex}})$ with $S_{ex}(t) = \frac{1}{2}\pi I_0 \gamma t^2$.

The calculations are performed for different values of the extended surface, $S_{ex}$, that implies different values of the fraction of surface, $S$, that is not available for actual nucleation: $S(t) = 1 - e^{-S_{ex}(t)}$ (see also Appendix B for computational details). The $S_{ex}(t)$ value gives the nucleation rate of actual nuclei at time $t$, since $\frac{I_a(t)}{I_0} = e^{-S_{ex}(t)}$. As stated above, in discussing the present results the phantom-included nucleation rate, $I_0 t$, is always assumed to be different from zero. In fact, for $I_0 t = 0$ no nuclei are present on the surface and the $g(r)$ quantity cannot be defined.

The $g(\rho, S_{ex})$ radial distribution functions are shown in Fig.2a for several values of $S_{ex}(t)$, where $\rho = \frac{r}{\sqrt{\gamma t}}$. In the limit of large $\rho$ these curves approach unity which is the correct behavior since, in this limit, correlation with the central nucleus vanishes. In Fig.2b the radial distribution functions are displayed as a function of $\tilde{\rho} = r\sqrt{2\pi N}$ namely, the reduced distance usually adopted in the literature for studying spatial distribution of nuclei at electrode surfaces [20]. Uncorrelated nucleation is attained for $S_{ex}(t) \to 0$. In fact, the computation at $S_{ex} = 10^{-3}$ provides $g \cong 1$ (Fig.2b). When plotted as a function of $\rho$, the uncorrelated case does not provide $g = 1$ in the whole $\rho$-domain of Fig.2a. This is explained by the relation $\tilde{\rho} = \rho\sqrt{4S_{ex}\phi_1(S_{ex})}$ that gives, at small $S_{ex}$, $\tilde{\rho} \approx \rho\sqrt{S_{ex}}$ and implies a strong "expansion" of the $\rho$ scale when compared to the $\tilde{\rho}$ one.

The nearest-neighbor probability densities are reported in Fig.3a as a function of $\tilde{\rho}$ and for the same set of $S$'s of Fig.2. The Poissonian distribution of actual nuclei is obtained in the limit of low $S$ values as shown by the curve at $S = 10^{-3}$ which is in excellent agreement with eqn.13. The symmetry of the nnd functions is quantified through the skwness parameter, namely $\gamma_1 = \frac{\mu_3}{\mu_2^{3/2}}$, where $\mu_i$ is the i$^{th}$ central moment of the distribution. The behavior of $\gamma_1$ as a function of $S$ for the nearest neighbor probability densities is displayed in the inset of Fig.3a. For low values of the exclusion area the skewness increases with respect to the value of the Poissonian distribution ($\gamma_1 = 0.63$), and then decreases to reach the asymptotic value $\gamma_1 = 0.44$ at saturation ($S = 0.99$). It follows that the highest symmetry of the nnd function is attained at saturation. Moreover, the mode of the distribution, $\tilde{\rho}_{max}$, increases monotonically with $S$ (inset of Fig.3a). In Fig.3b the distribution has been plotted as a function of the ratio $\rho_x = \frac{\tilde{\rho}}{\tilde{\rho}_{max}} = \frac{r}{r_{max}}$, a representation that will be used below for comparison with experimental data.

As anticipated in the introduction, analytical approaches of nnd have been proposed which are based on the hard-core interaction between disks equal in size. In Fig.4 the probability densities of Fig.3 are compared to that of the hard-disk model according to the theory developed by Torquato et al [32]. By denoting with $d_c$ the disk diameter, $\sigma_c = \pi N \frac{d_c^2}{4}$ is the fraction of surface covered by disks. The quantity $\sigma = 1 - e^{-4\sigma_c}$ has the same meaning as $S$ in the present modeling: it is the area of the region where nucleation is prevented. Fig.4 shows that the hard-disk model reproduces the curves only in the limit of small $S$, when the spread in size of the exclusion disks for nucleation is small. Conversely, at saturation the behavior of the two nnd's differ markedly. We notice, in passing, that the $H_p(\tilde{\rho})$'s of Fig.3 are well described by a modified stretched exponential function $H_p(\tilde{\rho}) = a\tilde{\rho}^m e^{-b\tilde{\rho}^p}$, with $a, b, m, n$ positive definite coefficients. At saturation, the values of the exponents are $m \cong 4.1$ and $p \cong 1.8$ to be compared to the random distribution $m = 1$ and $p = 2$.

Electrochemical nucleation is based on the concept of nucleation exclusion zone as it stems by the "planar diffusion zone approach" [13,14]. An exclusion disk for nucleation, of radius $r_d$, greater than nucleus radius, develops around each actual nucleus and grows according to a parabolic law. Application of the model discussed in section 2.2 to electrochemical nucleation therefore requires the identity $R(t - t') \equiv r_d(t - t') = \sqrt{\gamma(t - t')}$, that leads to eqn.15. Computer simulations of 2D nucleation with the development of exclusion zones have been performed by several authors with the purpose of studying spatial distribution of nuclei at electrode surfaces [20-22, 24-27, 36]. To check the validity of the present modeling we first consider the results by Scharifker et al [20] which made use of the same nucleation and growth laws employed in the present computation. In particular, parabolic growth of exclusion disks and constant nucleation rate of actual nuclei per unit surface available for nucleation. It is worth noting that this nucleation protocol is equivalent to set $I(t) = I_0$ since $\frac{I_a(t)}{(1-S)} = I_0$ is the nucleation rate of actual nuclei per unit of surface uncovered by exclusion zone, i.e. available for the nucleation of actual nuclei. In the simulation, the nnd has been determined after nucleation reached completion [20]. Comparison between the simulation and the analytical result, eqn.15, is displayed in Fig.5 as a function of the reduced distance $\tilde{\rho}$. The good agreement between theory and simulation indicates that the approximation above employed for the conditional radial distribution function is fairly good. This could be explained by the quite low values of the $g(\tilde{\rho})$ in a distance range that is of the order of magnitude of the width of the nearest neighbor probability density.

In Figs.6a-b and Figs.7a-c the computer simulations by Tsakova and Milchev [27] and by Guo and Searson [21] are reported as a function of $\rho_x = r/r_{max}$ together with the analytical results given by eqn.15. Even in this case the agreement between the simulation and the theory is satisfactory for saturation condition ($S$=0.99 and $S$=0.95 in Fig.6b and Fig.7c, respectively). Differences between the simulations of refs.[21,27] and both eqn.15 and the simulation of ref.[20] could be ascribed to the different nucleation mechanism. In particular, in the simulations reported in refs.[21,27] a "sharp" exclusion zone is not assumed and nucleation is allowed to occur in the planar diffusion zone, although with a reduced probability [18].

In Figs.6,7 the Poisson distribution is also reported which is in good agreement with the computer simulations of Figs.6a and Fig.7a. As discussed in refs.[21,27] this situation is related to the exhaustion of a limited number of seeds (active sites) involved in the nucleation process. This finding is in accord with the model presented here. In fact, according to eqn.15 the nnd depends upon $S_{ex} = \frac{1}{2}\pi I_0 \gamma t^2$ that can be rewritten as $S_{ex} = \frac{1}{2}\pi N_p(t)R(t)^2$, where $N_p(t)$ is the phantom-included nucleation density, and $R(t)$ the maximum nucleus size. The Possonian nnd is therefore recovered either for small nucleation densities ($N_p(t)$) or low growth rate of the nuclei. The former case should pertain to the simulations of Fig.6a and Fig.7a. Therefore, the analytical model does show that the transition from Poissonian to non-Poissonian nnd is ruled by the single parameter $S_{ex}$ (or even $S$).

Finally, in Figs.8 and 9 the numerical solutions of eqn.15 have been compared with experimental data on electrochemical nucleation of Mercury on Platinum [22] and of Silver on Boron-doped diamond electrodes [25]. Since in ref.[22] the experimental nnd was expressed in terms of the $r$ distance, in Fig.8 the probability distribution has been reported as a function of the reduced distance $\rho_x = r/r_{max}$ The best agreement between experimental data and eqn.15 is attained for $S \cong 0.1$ and $S = 0.95$ for the deposition of Hg on Pt (Fig.8a,b) and for $S = 0.4$ for Ag on doped diamond (Fig.9). Notably, for the data of Fig.8b the value S=0.95 is close to saturation condition in accord with author's results [22]. On the other hand, data points described by the curve at S=0.1 (panel a in Fig.8), which are closer to the Poisson nnd, refers to the earlier stage of the electrodeposition [22]. It stems that the limiting case of progressive nucleation is capable of reproducing the salient features of the distribution including the transition from Poissonian to correlated nearest neighbor probability density.

### 3- Conclusions

Phase transformations ruled by non-simultaneous nucleation lead to non-random distribution of actual nuclei. In this paper an analytical approach has been developed for modeling the pair distribution function of actual nuclei in the case of progressive nucleation. On the basis of the approach proposed in ref.[32] the pair distribution function is further used to estimate the nearest neighbor spatial distribution. It is shown that the nnd function depends on reduced distance and extended surface of the region where nucleation is prevented. Such a region can either coincide with the new phase or to be of greater extension, depending on nucleation mechanism. At short time the nnd is shown to be nearly Poissonian, with skewness $\gamma_1 = 0.63$, and evolves to non-Poissonian distributions to reach a $\gamma_1 = 0.44$ at saturation. The present modeling finds application also in the field of electrochemical nucleation, under diffusion controlled conditions, where correlation effects among nuclei are significant. The analytical model is shown to be in good agreement with both computer simulations and experimental results of nnd on electrochemical nucleation.

**Appendix A**

For $R(t) = (\gamma t)^{n/2}$ the integrands of eqns.2a-b become ($t' < t'' < t$):

$$A_1(t', \tau) = \pi \gamma^n (t' - \tau)^n \tag{A1}$$

$$A_2(t', t'', \tau, r) = \pi \gamma^n (t'' - \tau)^n - \omega_1[\gamma^{n/2}(t' - \tau)^{n/2}, \gamma^{n/2}(t'' - \tau)^{n/2}; r], \tag{A2}$$

$$A_3(t', t'', \tau, r) = \pi \gamma^n (t'' - \tau)^n - \omega_2[\gamma^{n/2}(\tau - t')^{n/2}, \gamma^{n/2}(t'' - \tau)^{n/2}; r], \tag{A3}$$

where

$$\omega[R_1, R_2; r] = R_2^2 \cos^{-1} \frac{r^2 - (R_1^2 - R_2^2)}{2rR_2} + R_1^2 \cos^{-1} \frac{r^2 + (R_1^2 - R_2^2)}{2rR_1} - \frac{1}{2}[4r^2 R_2^2 - (r^2 - R_1^2 + R_2^2)^2]^{1/2} \tag{A4}$$

The argument of the square root in eqn.A4 is positive definite for $|R_1 - R_2| < r < R_1 + R_2$. Using dimensionless variables, $x' = t'/t$, $x'' = t''/t$, $\eta = \tau/t$, $\rho = r/(\gamma t)^{n/2}$ and $\bar{\omega}_i = \omega_i/(\gamma t)^n$, one obtains ($x' < x'' < 1$):

$$\int_0^{t'} [A_1(t', \tau) + A_2(t', t'', \tau, r)] I(\tau) \, d\tau$$

$$= \pi I_0 \gamma^n t^{n+1} \int_0^{x'} [(x' - \eta)^n + (x'' - \eta)^n - \frac{1}{\pi} \bar{\omega}_1(x', x'', \eta, \rho)] d\eta$$

$$= \frac{1}{n+1} \pi I_0 \gamma^n t^{n+1} \left[ (x')^{n+1} + (x'')^{n+1} - (x'' - x')^{n+1} \right.$$

$$\left. - \frac{n+1}{\pi} \int_0^{x'} \bar{\omega}_1(x', x'', \eta, \rho) \, d\eta \right], \tag{A5}$$

where

$$\overline{\omega}_1(x', x'', \eta, \rho) = (x'' - \eta)^n \cos^{-1} \frac{\rho^2 - [(x'-\eta)^n - (x''-\eta)^n]}{2\rho(x''-\eta)^{n/2}} + (x' - \eta)^n \cos^{-1} \frac{\rho^2 + [(x'-\eta)^n - (x''-\eta)^n]}{2\rho(x'-\eta)^{n/2}} -$$
$$\frac{1}{2}\sqrt{2\rho^2[(x'-\eta)^n + (x''-\eta)^n] - \rho^4 - [(x'-\eta)^n - (x''-\eta)^n]^2} \quad .$$

For the last integral in eqn.2b one obtains:

$$\int_{t'}^{t''} A_3(t', t'', \tau, r) I(\tau) d\tau = \pi I_0 \gamma^n t^{n+1} \int_{x'}^{x''} \left[ (x'' - \eta)^n - \frac{1}{\pi}\overline{\omega}_2(x', x'', \eta, \rho) \right] d\eta =$$
$$\frac{1}{n+1} \pi I_0 \gamma^n t^{n+1} \left[ (x'' - x')^{n+1} - \frac{n+1}{\pi} \int_{x'}^{x''} \overline{\omega}_2(x', x'', \eta, \rho) \, d\eta \right]. \tag{A6}$$

where

$$\overline{\omega}_2(x', x'', \eta, \rho) = (x'' - \eta)^n \cos^{-1} \frac{\rho^2 - [(\eta-x')^n - (x''-\eta)^n]}{2\rho(x''-\eta)^{n/2}} + (\eta - x')^n \cos^{-1} \frac{\rho^2 - [(x''-\eta)^n(\eta-x')^n]}{2\rho(\eta-x')^{n/2}} -$$
$$\frac{1}{2}\sqrt{2\rho^2[(x''-\eta)^n + (\eta-x')^n] - \rho^4 - [(\eta-x')^n - (x''-\eta)^n]^2}$$

The function $P(t, x'', x', \rho)$ eventually becomes:

$$P(t, x'', x', \rho) =$$
$$\exp\left[ -\frac{1}{n+1}\pi I_0 \gamma^n t^{n+1} \left[ (x')^{n+1} + (x'')^{n+1} - (x''-x')^{n+1} - \frac{n+1}{\pi} \int_0^{x'} \overline{\omega}_1(x', x'', \eta, \rho) \, d\eta \right] -$$
$$\frac{1}{n+1}\pi I_0 \gamma^n t^{n+1} \left[ (x'' - x')^{n+1} - \frac{n+1}{\pi} \int_{x'}^{x''} \overline{\omega}_2(x', x'', \eta, \rho) \, d\eta \right] \right]$$

$$= \exp\left[ \frac{1}{\pi}(n+1)S_{ex}(t) \left[ \int_0^{x'} \overline{\omega}_1(x', x'', \eta, \rho) \, d\eta + \int_{x'}^{x''} \overline{\omega}_2(x', x'', \eta, \rho) \, d\eta \right] - S_{ex}(t)(x''^{n+1} + x'^{n+1}) \right] , \tag{A7}$$

where $S_{ex}(t) = \frac{1}{n+1}\pi I_0 \gamma^n t^{n+1}$ is the extended surface of the exclusion zones for nucleation. Finally, use of eqn.A7 in eqn.5b provides the $f$-function eqn.6.

**Appendix B**

Let us consider the integral in the numerator of eqn.7, namely

$$\int_0^1 e^{-S_{ex}x''^{n+1}}dx'' \int_0^{x''} e^{-S_{ex}x'^{n+1}} e^{\frac{n+1}{\pi}S_{ex}\,\Omega(x',x'',\rho)} H\!\left(x'-x''+\rho^{2/n}\right)dx',$$

where $\Omega(x',x'',\rho)=\int_0^{x'}\overline{\omega}_1(x',x'',\eta,\rho)\,d\eta+\int_{x'}^{x''}\overline{\omega}_2(x',x'',\eta,\rho)\,d\eta$ and $S_{ex}\equiv S_{ex}(t)$. Since $x'>x''-\rho^{2/n}$, with $x'>0$, the integral becomes:

$$\left[\int_0^{\rho^{2/n}} e^{-S_{ex}x''^{n+1}}dx'' \int_0^{x''} e^{-S_{ex}x'^{n+1}} e^{\frac{n+1}{\pi}S_{ex}\,\Omega(x',x'',\rho)}dx'\right.$$

$$\left.+\int_{\rho^{2/n}}^1 e^{-S_{ex}x''^{n+1}}dx'' \int_{x''-\rho^{2/n}}^{x''} e^{-S_{ex}x'^{n+1}} e^{\frac{n+1}{\pi}S_{ex}\,\Omega(x',x'',\rho)}dx'\right] H(1-\rho)$$

$$+\left[\int_0^1 e^{-S_{ex}x''^{n+1}}dx'' \int_0^{x''} e^{-S_{ex}x'^{n+1}} e^{\frac{n+1}{\pi}S_{ex}\,\Omega(x',x'',\rho)}dx'\right] H(\rho-1). \qquad (A8)$$

Changing the order of integration in the first bracket of eqn.A8 and setting $n=1$, one eventually obtains

$$\left[\int_0^{1-\rho^2} e^{-S_{ex}x'^2}dx' \int_{x'}^{x'+\rho^2} e^{-S_{ex}x''^2} e^{\frac{2}{\pi}S_{ex}\,\Omega(x',x'',\rho)}dx''\right.$$

$$\left.+\int_{1-\rho^2}^1 e^{-S_{ex}x'^2}dx' \int_{x'}^1 e^{-S_{ex}x''^2} e^{\frac{2}{\pi}S_{ex}\,\Omega(x',x'',\rho)}dx''\right] H(1-\rho)$$

$$+\left[\int_0^1 e^{-S_{ex}x''^2}dx'' \int_0^{x''} e^{-S_{ex}x'^2} e^{\frac{2}{\pi}S_{ex}\,\Omega(x',x'',\rho)}dx'\right] H(\rho-1)\;, \qquad (A9)$$

that is the expression employed in numerator of eqns.14,15 to compute, numerically, the $g(\rho, S_{ex})$ and the nnd functions.

**Figure captions**

Fig.1 Pictorial view of the areas (capture zones) $A_1$, $A_2$ $A_3$ and $\omega_i$ defined in eqns.2a-b.

Panels a) and b) refer to time sequence $\tau < t' < t'' < t$, for $r < R(t' - \tau) + R(t'' - \tau)$ and $r > R(t' - \tau) + R(t'' - \tau)$, respectively. The two seeds at relative distance $r$ do not belong to the exclusion zone for nucleation, within time $t'$, provided that no nucleation event takes place, between $d\tau$ at time $\tau$, in the region enclosed by the thick dashed line. The area of this region is equal to $A_1 + A_2$.

Panels c) and d) illustrate the cases $t' < \tau < t'' < t$ and $t' < t'' < \tau < t$, respectively. In c) the second seed do not belong to the exclusion zone for nucleation, within time $t''$, provided that no nucleation event takes place, between $d\tau$ at $\tau$, in the colored region of area $A_3$. The area $\omega_2$ is different from zero for $r < R(\tau - t') + R(t'' - \tau)$. In panel d) the area to be considered is nil.

Fig.2. Pair distribution function as a function of reduced distance $\rho = r/(\gamma t)^{1/2}$ and $\tilde{\rho} = r\sqrt{2\pi N}$ are reported in panels a) and b), respectively, for several values of the surface where nucleation is prevented ($S$). $S$ values are reported in the figure and increase from one curve to the other according to the direction of the arrow : a) S=0.095, 0.5, 0.86, 0.99; b) S=10$^{-3}$, 0.095, 0.4, 0.5, 0.86, 0.95, 0.99.

Fig.3 Nearest neighbor probability density computed from eqn.15, are reported in panels a) and b) as a function of $\tilde{\rho} = r\sqrt{2\pi N}$ and $\rho_x = \frac{\tilde{\rho}}{\tilde{\rho}_{max}} = \frac{r}{r_{max}}$, respectively. The computation refers to several values of the surface where nucleation is prevented ($S$). $S$ values are reported in the panels and increase from one curve to the other according to the direction of the arrows: S=10$^{-3}$, 0.095, 0.26, 0.4, 0.5, 0.86, 0.95, 0.99. In panel a) solid symbols highlight the computation at $S = 10^{-3}$ that is in excellent agreement with Poisson distribution (thin line through points). In the inset of panel a) the behavior of both the skweness parameter ($\gamma_1$, red points) and the mode of distribution ($\tilde{\rho}_{max}$, blue points) are displayed. Dashed line in panel b) is the Poisson nnd.

Fig.4. Comparison between the nnd of the present approach (solid lines) and those of the hard-disk model according to ref.[32] (dashed lines). The values of the exclusion areas for hard-core correlation, $\sigma$, are reported in the figure. For the nnd function a, b and c, the $S$ values are 0.1, 0.4 and 0.95, respectively.

Fig.5 The nearest neighbor probability density given by the analytical model (eqn.15) is compared to the computer simulation of ref.[20]. The reduced distance is $\tilde{\rho} = r\sqrt{2\pi N(t)}$. The differences between the two nnd are due to the approximation of the $G(r)$ and to the finite number of nuclei in the simulation.

Fig.6 Comparison between computer simulations of nnd performed in [27] and eqn.15. Panels a), b) refer to different nucleation densities which is found to affect the degree of correlation among actual nuclei. The distribution in panel a) is nearly Poissonian which implies, according to eqn.15, small $S$ values . In panel b) a $S$=0.99 is used to better fit the histogram; this implies completion of the nucleation process. In panels a) and b), the Poisson nnd is shown as solid and dashed lines respectively.

Fig.7 Comparison between computer simulations of nnd performed in [21] and eqn.15. Panels a), b) and c) refer to different nucleation densities i.e. increasing time of deposition. The distribution in panel a is nearly Poissonian which implies, according to eqn.15, small $S$ values. Higher $S$ values are used to better fit the histograms of panels b) and c), where Poisson nnd is shown as dashed line.

Fig.8) Experimental probability densities for Hg on Pt according to ref. [22]. The best agreement between the analytical model and the experimental data is attained for $S$=0.095 (panel a) and $S$=0.95 (panel b). Dashed line is the Poisson nnd.

Fig.9) Experimental probability densities for Ag on Boron doped diamond electrode [25]. The analytical computation (eqn.15) for $S$=0.4 is displayed as solid line. Black dashed line is the Poisson nnd.

**Figures**

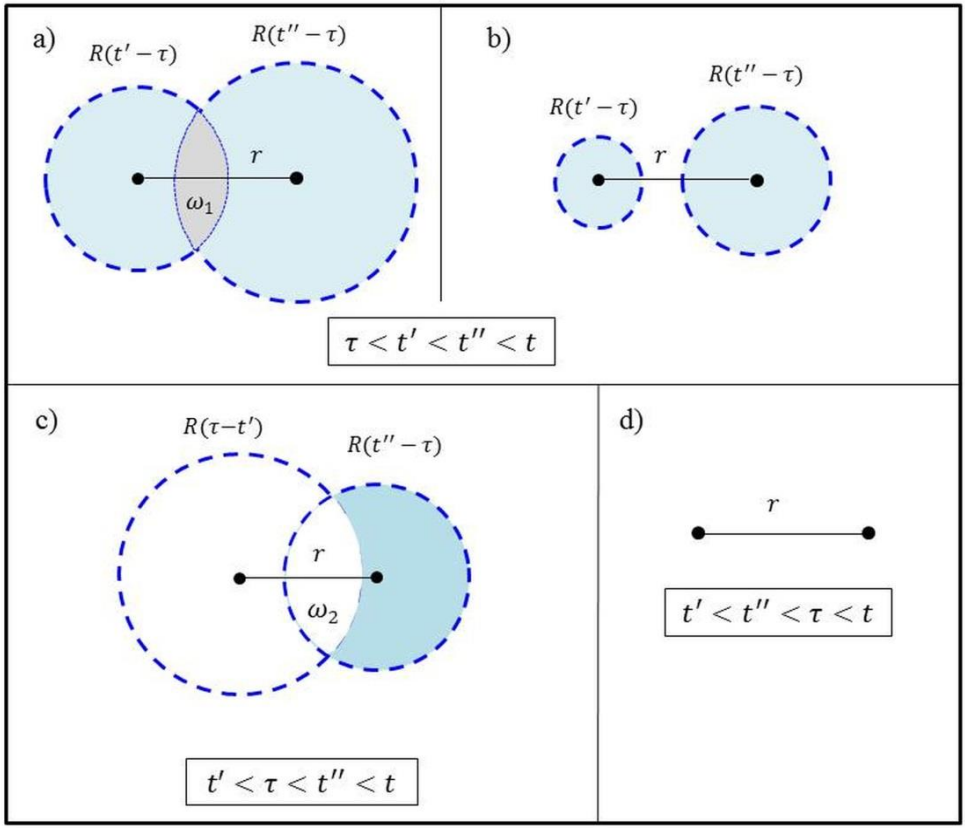

Fig.1

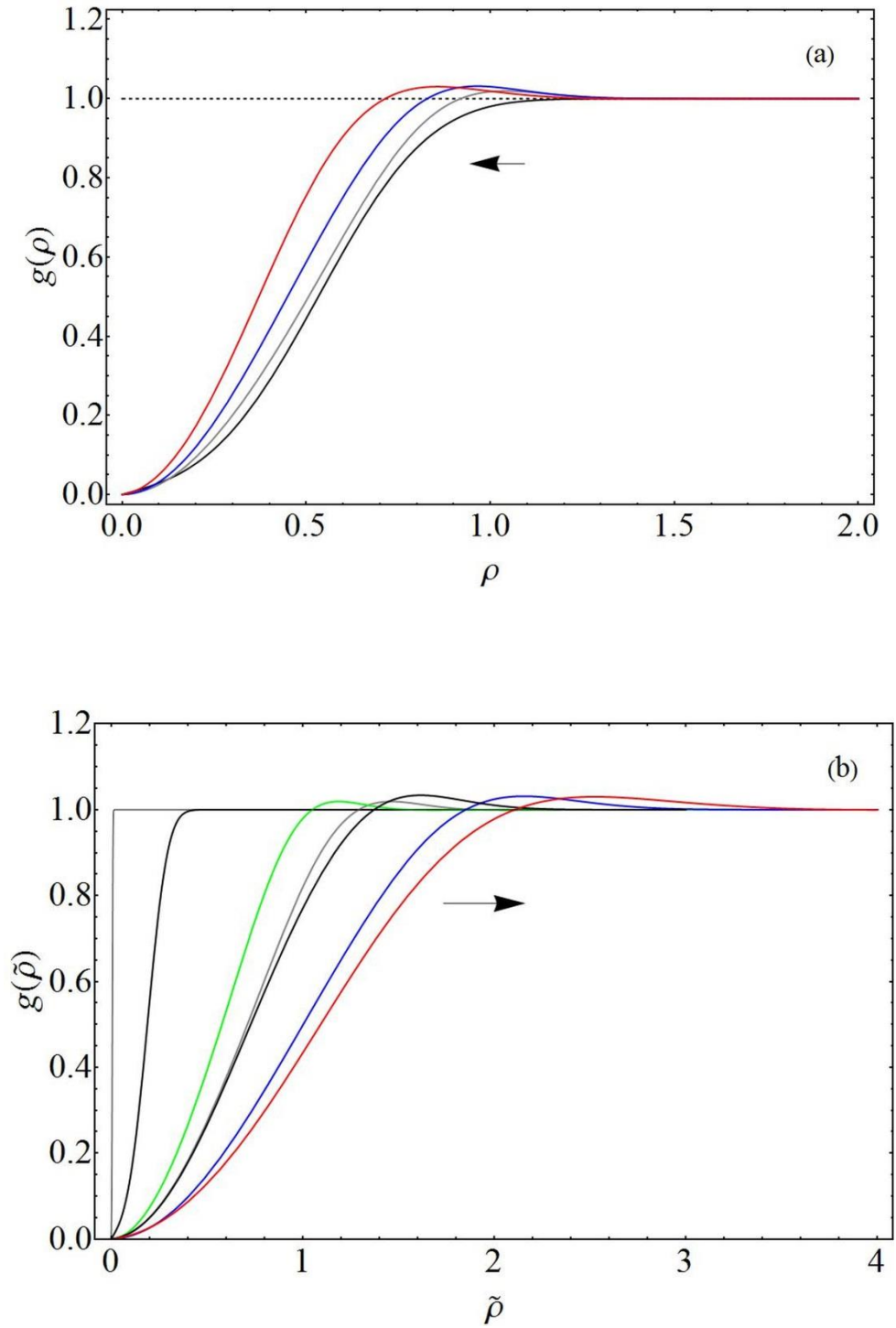

Fig.2

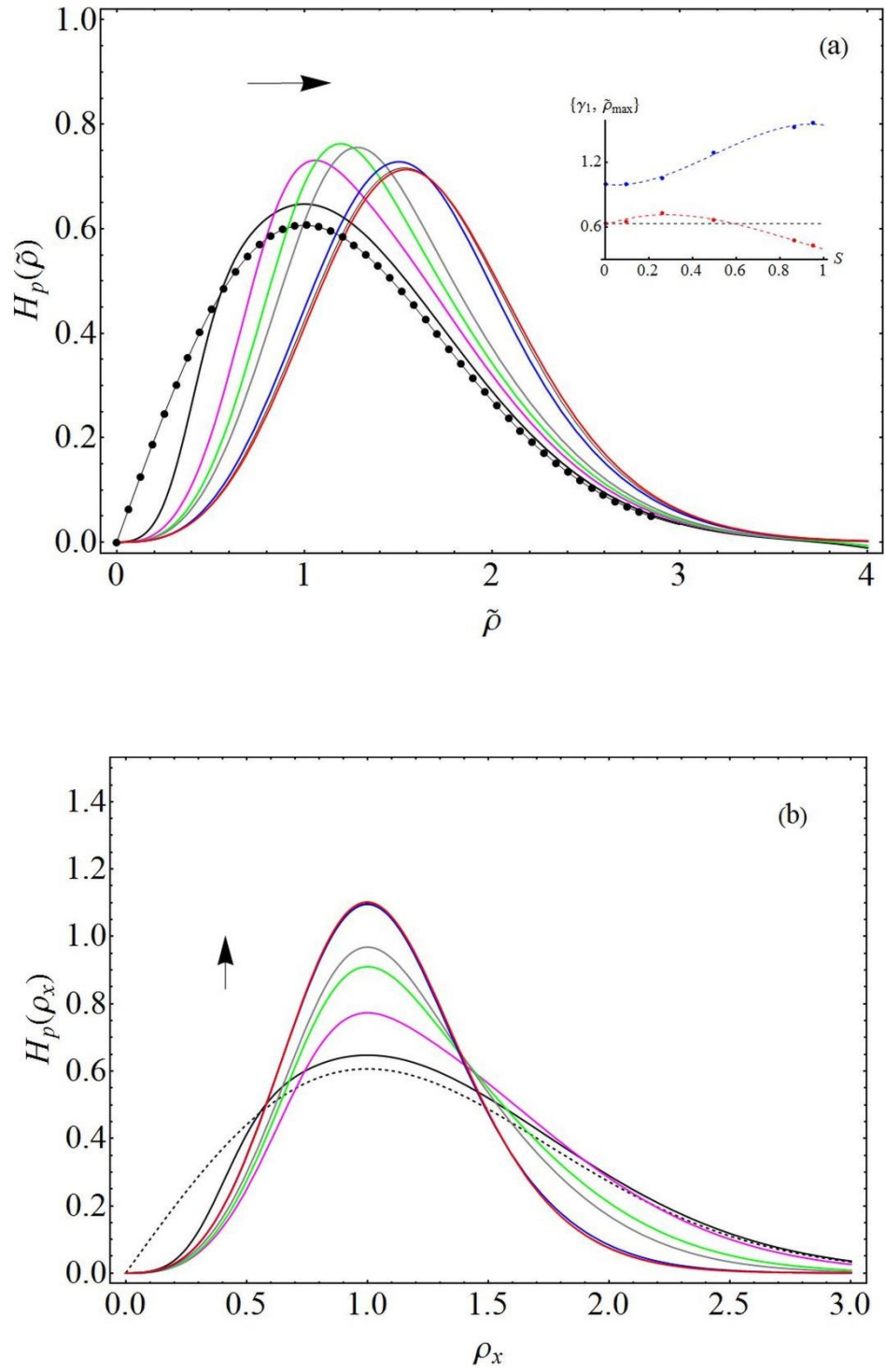

Fig.3

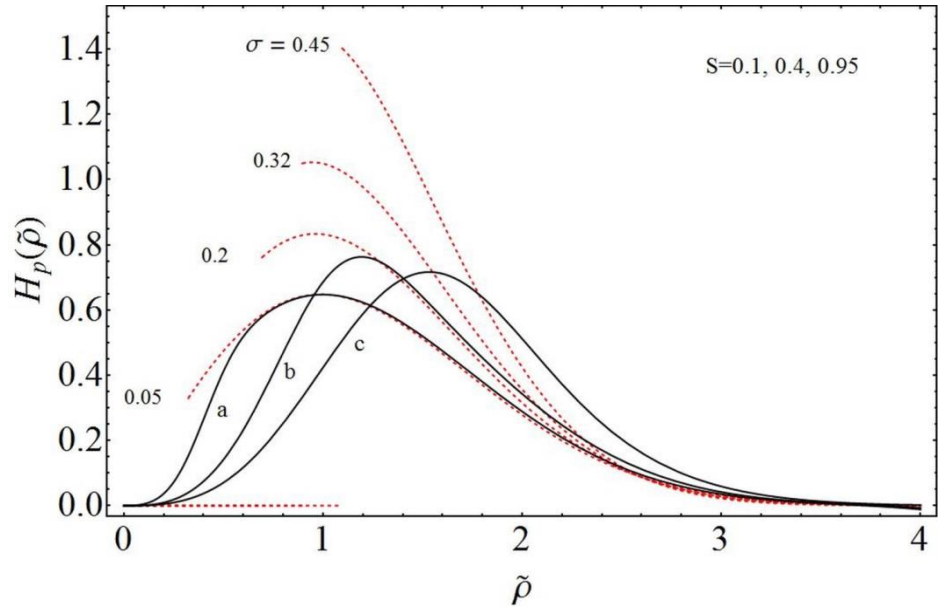

Fig.4

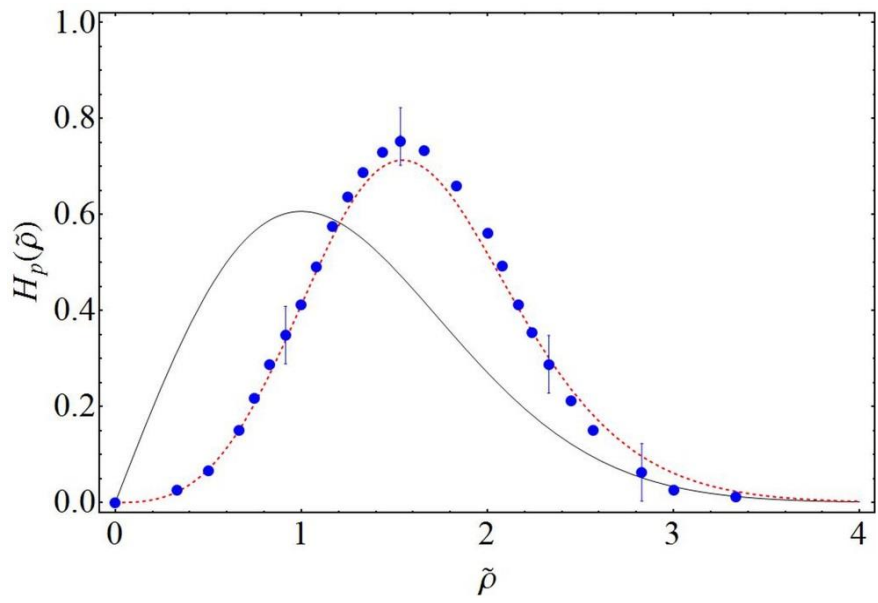

Fig.5

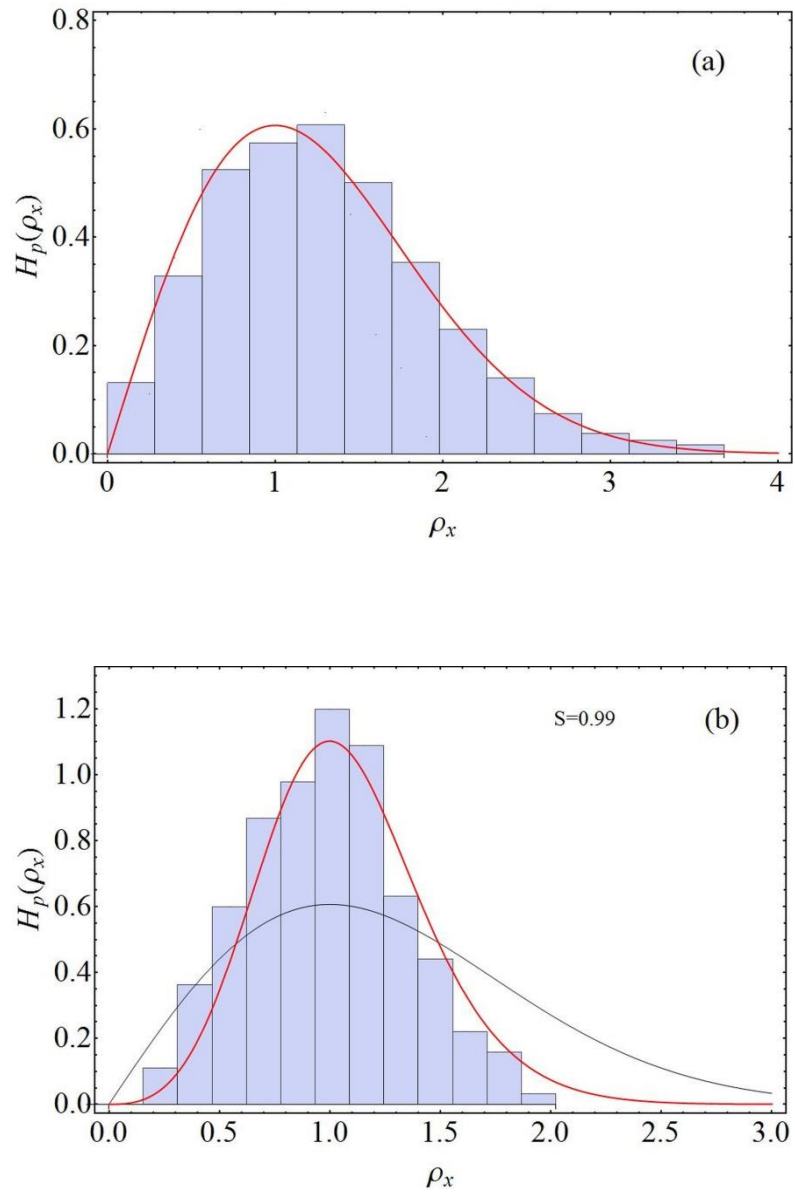

Fig.6

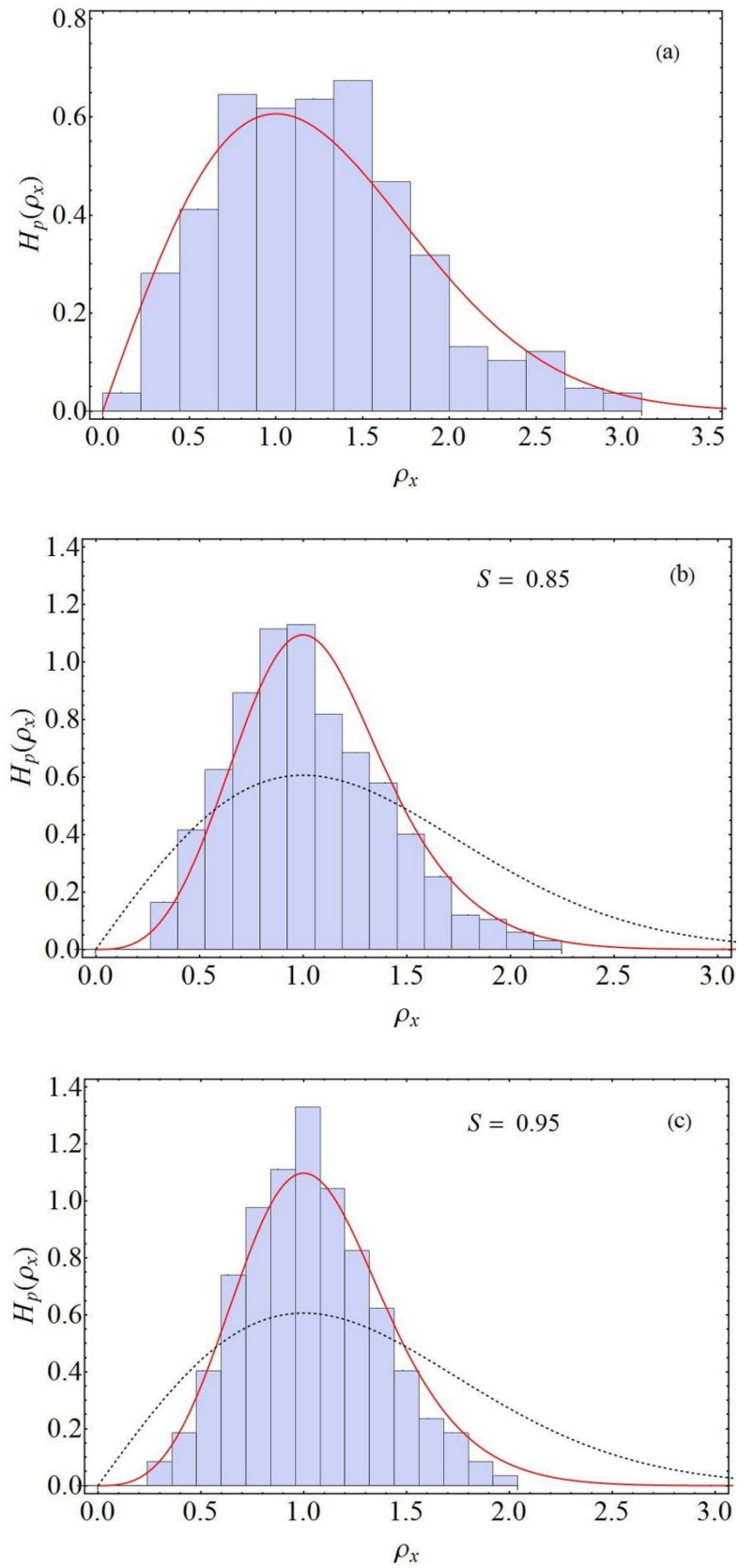

Fig.7

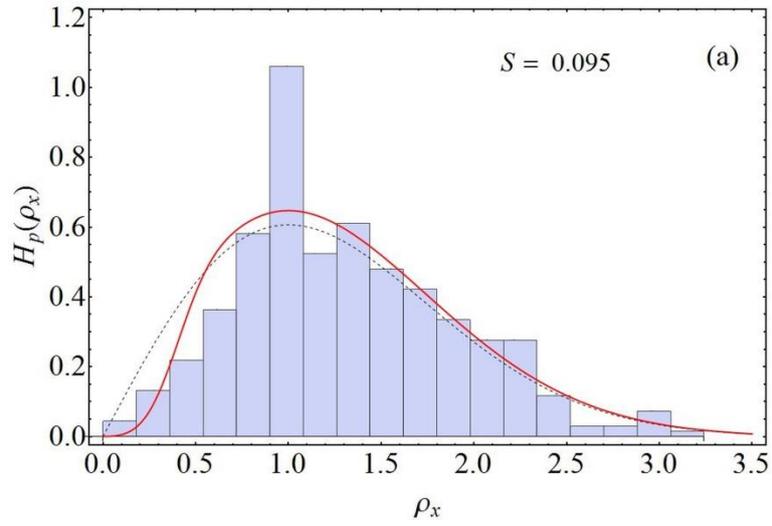

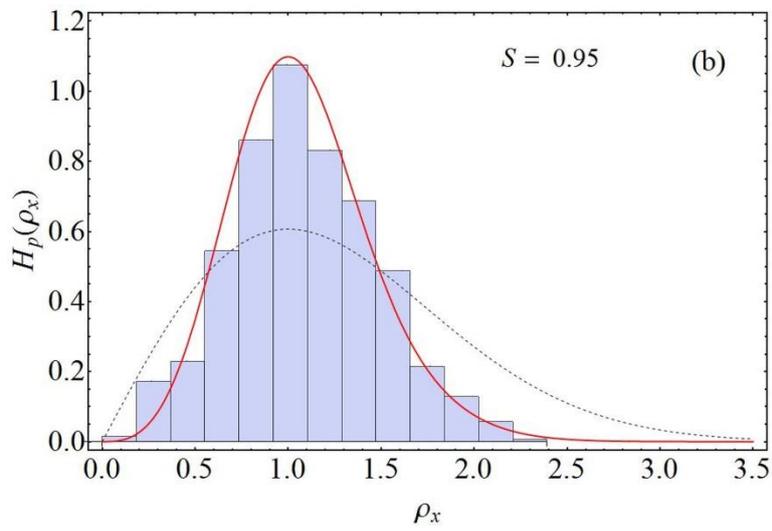

Fig.8

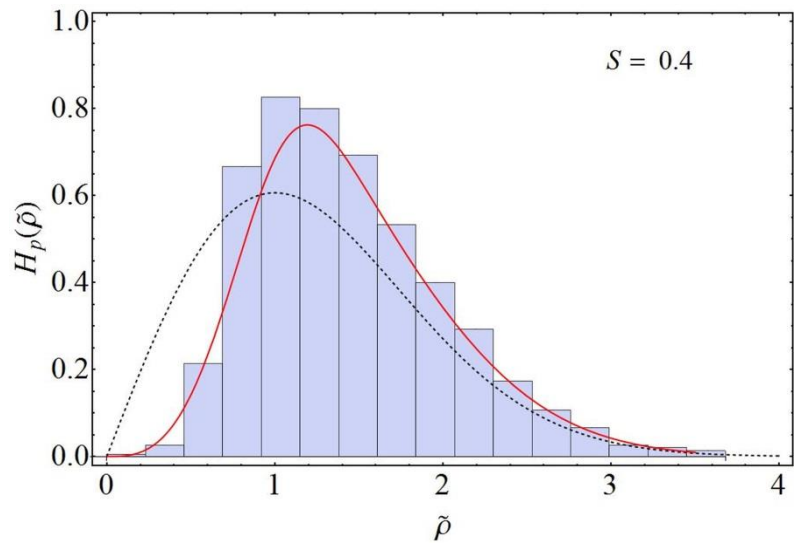

Fig.9